# Production of Ultra-Thin and High-Quality Nanosheet Networks via Layer-by-Layer Assembly at Liquid-Liquid Interfaces

Joseph Neilson[1], Eoin Caffrey[1], Oran Cassidy[1], Cian Gabbett,[1] Kevin Synnatchke[2], Eileen Schneider[3], Jose M. Munuera[4], Tian Carey,[1] Max Rimmer[5], Zdenek Sofer[6], Janina Maultzsch[3], Sarah J. Haigh[5] and Jonathan N. Coleman[1]*

[1]*School of Physics, CRANN & AMBER Research Centres, Trinity College Dublin, Dublin 2, Ireland*

[2]*Dresden University of Technology, Dresden, Germany*

[3]*Department of Physics, Friedrich-Alexander-Universität Erlangen-Nürnberg, Staudtstr. 7, 91058 Erlangen, Germany*

[4] *Department of Physics, Faculty of Sciences, University of Oviedo, C/ Leopoldo Calvo Sotelo, 18, 33007 Oviedo, Asturias, Spain.*

[5] *Department of Materials and National Graphene Institute, The University of Manchester, Oxford Rd, Manchester, M13 9PL, UK*

[6] *Department of Inorganic Chemistry, University of Chemistry and Technology Prague, Technická 5, 166 28 Prague 6, Czech Republic*

*colemaj@tcd.ie (Jonathan N. Coleman); Tel: +353 (0) 1 8963859.

ABSTRACT

Solution-processable 2D materials are promising candidates for a range of printed electronics applications. Yet maximising their potential requires solution-phase processing of nanosheets into high-quality networks with carrier mobility ($\mu_{Net}$) as close as possible to that of individual nanosheets ($\mu_{NS}$). In practise, the presence of inter-nanosheet junctions generally limits electronic conduction, such that the ratio of junction resistance ($R_J$) to nanosheet resistance ($R_{NS}$), determines the network mobility via $\mu_{NS} / \mu_{Net} \approx R_J / R_{NS} + 1$. Hence, achieving $R_J/R_{NS}<1$ is a crucial step for implementation of 2D materials in printed electronics applications. In this work, we utilise an advanced liquid-interface deposition process to maximise nanosheet alignment and network uniformity, thus reducing $R_J$. We demonstrate the approach using graphene and MoS$_2$ as model materials, achieving low $R_J/R_{NS}$ values of 0.5 and 0.2, respectively. The resultant graphene networks show a high conductivity of $\sigma_{Net} = 5 \times 10^4$ S/m while our semiconducting MoS$_2$ networks demonstrate record mobility of $\mu_{Net}$ = 30

cm$^2$/Vs, both at extremely low network thickness (t$_{Net}$ <10 nm). Finally, we show that the deposition process is compatible with non-layered quasi-2D materials such as silver nanosheets (AgNS), achieving network conductivity close to bulk silver for networks <100 nm thick. We believe this work is the first to report nanosheet networks with R$_J$/R$_{NS}$<1 and serves to guide future work in 2D materials-based printed electronics.

INTRODUCTION

The field of printed electronics is rapidly expanding with solution processing facilitating the fabrication of a wide range of electronic components.[1, 2] While printed electronic devices tend to display reduced performance compared to traditional silicon-based electronics, they have considerable advantages like low-cost, mechanical flexibility, and large area deposition compatibility.[1] As such, they could find future applications in wearable devices, point-of-care healthcare applications, e-skins and many more. Possibly the most important classes of materials for printed electronics have been organic molecules and polymers. However, although organics have been very successful due to their processability and versatility, organic devices are starting to reach performance limits. For example, maximum device mobilities have remained around 10 $cm^2$/Vs for a number of years.[3]

One strategy to address this limitation has been to turn to inorganic nanomaterials as the active materials in printed devices.[4] Of the various classes of nanomaterials being utilised in this area, 2D materials are particularly exciting.[5] 2D materials are characterised by nanometer-scale thickness, and large aspect ratio (ratio of lateral size to thickness).[6] Crucially, the family of 2D materials is extremely broad, containing thousands of members with a wide range of properties.[7] It contains conductors such as graphene and semiconductors such as molybdenum disulphide ($MoS_2$), both types displaying relatively high intrinsic mobility, as well as insulators such as hBN or BiOCl, which display high breakdown strength or permittivity.[8, 9]

Critically, 2D materials can be dispersed in liquids using a range of methods to form high quality inks that can be solution-deposited using various techniques.[10-14] Their wide range of electronic characteristics, inherent flexibility, and compatibility with solution-based processing, makes 2D materials the ideal functional materials for future printed electronics inks.[15-17] These inks have been used to deposit nanosheets and networks for a range of device applications including non-volatile memory,[18] memristors,[19] transistors,[20-23] solar cells[24] and photodiodes.[25, 26]

However, to exploit 2D materials in these applications, it will be necessary to optimise methods to solution-deposit high-quality networks over large areas. It is essential that these printed networks must display properties suitable for use in devices, such as high mobility or conductivity.

A drawback of printing of inks of nanosheets with low aspect ratios (< roughly 50) is that it often results in networks that are extremely porous and disordered.[27] As a result, the junctions between nanosheets, which are known to limit network mobility and conductivity in most cases, tend to be point-like and display extremely high junction resistances ($R_J$).[28] These very high junction resistances result in low network conductivity and mobility. For example, it has been shown[28] that printed networks of $WS_2$ nanosheets produced by liquid phase exfoliation have junction resistances of 25 G$\Omega$, a value which is 100-1000 higher (depending on nanosheet dimensions) than the resistance of the $WS_2$ nanosheets themselves ($R_{NS}$). The ratio of junction to nanosheet resistance ($R_J/R_{NS}$) is a critical parameter for determining the electrical properties of networks.[28] Ideally, $R_J < R_{NS}$ should lead to network properties approaching those of the nanosheets themselves. On the other hand, values of $R_J/R_{NS} >> 1$, will lead to network conductivity and mobility well below intrinsic nanosheet values.

Significant progress has been made towards addressing this problem in recent years. It has been found that nanosheets with large aspect ratios (>>100, usually made by electrochemical exfoliation[10]) form conformal junctions,[5] resulting in low-porosity networks consisting of highly aligned nanosheets.[22] Such alignment appears to yield networks with high conductivities in conducting films (up to $10^5$ S/m in graphene[5] and up to $10^6$ S/m in MXenes[29]) and large mobilities in semiconducting films (up to 11 cm$^2$/Vs in $MoS_2$).[20, 22] Recent results have shown that these high conductivities/mobilities can be linked to relatively low junction resistances.[28] For example, impedance spectroscopy has revealed relatively low junction resistances of $R_J \sim$ 1 M$\Omega$[28] in highly-aligned networks of high-aspect-ratio electrochemically exfoliated $MoS_2$ compared to $R_J$ = 25 G$\Omega$[28] for poorly aligned[27] networks of low aspect ratio $WS_2$.

Despite this progress, the network mobilities for the $MoS_2$ networks described above[28] was only $\mu_{Net}$ ~7 cm$^2$/Vs, much lower than that of the nanosheets themselves ($\mu_{NS}$~40 cm$^2$/Vs)[28], corresponding to $R_J/R_{NS}$ ~6 and $\mu_{Net}/\mu_{NS}$~0.17. Therefore, to produce nanosheet networks with mobilities approaching the nanosheets themselves (i.e. $\mu_{NS} \sim \mu_{Net}$), requires that junction resistances are reduced further so that $R_J<R_{NS}$. While various ideas have been proposed to achieve this, e.g. chemical cross-linking,[30, 31] we believe that a fruitful approach is to gain more control over the deposition process, leading to better control of junctions and further reductions in junction resistance.

Here, we report the deposition of large-area, densely tiled nanosheet networks, via assembly at an immiscible liquid-liquid interface. A single deposition process yields a highly aligned monolayer of edge connected individual nanosheets. Subsequent depositions can then build up highly aligned laminar multilayers. This is in contrast to spray or spin coating processes whereby the random coverage is typically determined by the Poisson distribution.[32] Further, the liquid-liquid deposition used here stands out from Langmuir-Blodgett and other layer-by-layer assembly processes for its reduced complexity in terms of equipment and ink formulation, respectively. In Langmuir-Blodgett deposition, a barrier is required to compress the floating monolayer.[33] In contrast, our monolayers are densified at the interface by Marangoni convection – a result of injecting the 2D ink at the interface. Alternatively, in layer-by-layer assembly, inks with oppositely charged zeta potential must be deposited in sequence to build networks with alternating positive and negatively charged layers.[34] The liquid-liquid interface assembly here does not require charged materials.

Here, we demonstrate the capability of the deposition process to repeatedly deposit monolayers of conducting (graphene) and semiconducting ($MoS_2$) nanosheets, as well as a metallic quasi-2D material (silver nanosheets) to form aligned, multilayer networks. We systematically measure the thickness-dependant optoelectrical properties of these networks and, for the first time, report networks displaying $R_J < R_{NS}$. The ability to deposit a range of 2D materials in this way opens avenues toward next-generation 2D materials-based heterostructures.

RESULTS AND DISCUSSION

*2D Materials Inks*

Throughout this work, we make use of three different nanomaterials: electrochemically exfoliated graphene, electrochemically exfoliated $MoS_2$, and quasi-2D colloidal Ag nanosheets[35]. The production of isopropyl-based inks from these materials is outlined in Methods. We characterised the dimensions of the nanosheets within the inks by performing AFM on nanosheets drop cast onto $Si/SiO_2$ substrates. Representative AFM images of the nanosheets are shown in figure 1A-C. Data for nanosheet length (L), width (W) and thickness (t) were extracted from these images and plotted in figure 1 D-F as nanosheet lateral size, represented as $(LW)^{1/2}$, plotted versus nanosheet thickness. The graphene, $MoS_2$ and AgNS nanosheets have mean nanosheet thicknesses of 1.3 ±0.1 nm, 0.6 ±0.02 nm, and 38 ±1 nm, lateral sizes of 2160 ±100 nm, 1170 ±65 nm, 560 ±12 nm and aspect ratios of 1850 ±90, 2010 ±130, and 17 ±1, respectively (all errors are standard errors).

*Liquid-liquid Interface Deposition*

We employ a deposition process that utilises the assembly of 2D materials at the immiscible water-hexane interface (Figure 2A),[32] applying it to our graphene, $MoS_2$ and AgNS inks. The liquid interface deposition of these materials results in densely tiled monolayers over a large area as illustrated via AFM images (Figure 2 B-D) and SEM images (Figure 2E-G). The ability of the liquid-liquid interface deposition process to fabricate such densely tiled monolayers can be attributed to the following factors. First, the thermodynamic stability of a nanosheet trapped at the high-energy water-hexane interface is high: the energy required to detach a nanosheet from the interface can be as large as $10^7$ $k_B T/\mu m^2$ for an $MoS_2$ nanosheet.[32] This ensures that nanosheets are tightly confined to the interface with minimal tendency to overlap. Next, the injection of IPA ink at the interface results in a decreased interfacial tension at the point of injection. The resulting interfacial tension gradient eventually acts to compress nanosheets into a jammed monolayer with predominantly edge-to-edge nanosheet contact. Finally, inter-nanosheet capillary forces may act to stabilise the interface-assembled monolayer as it is deposited onto a substrate.[36]

Next, we utilised a layer-by-layer assembly approach to fabricate thicker networks, enabling us to explore their optoelectrical properties as a function of thickness later in this work. We refer to a network consisting of one layer as "1L", two layers "2L" and so on. Due to uncertainty in thickness measurement for such thin networks, mean network thickness was found by averaging over several measurement methods (e.g. atomic force microscopy, white light interferometry, transmission electron microscopy) as outlined in Methods. The repeated layering of graphene, $MoS_2$ and Ag monolayers results in a linear trend in network thickness as a function of the number of deposited layers, with slopes of 2.38 ±0.07, 1.49 ±0.02, and 41.6 ±3.1 nm/layer, respectively (Figure 2 H-J). These values are in reasonable agreement with the mean nanosheet thicknesses in the inks, showing that each layer is close to a monolayer of the nanosheets present in the ink, with the layer thickness controlled by the average thickness of those nanosheets. A cross sectional TEM micrograph of a 5-layer $MoS_2$ network is shown in Figure 2K, and clearly shows a well-aligned network morphology (See also Figure S1). N.B. $MoS_2$ was imaged in this way as a model system due to improved contrast and ease of processing in comparison to graphene.

*Graphene Stacked Monolayers – Optoelectrical Properties*

The amount of light absorbed by a pristine atomic layer of graphene is well known to be $\pi e^2 / \hbar c = 2.3$ % per monolayer, where $c$, $e$, and $\hbar$ are the speed of light, electron charge, and reduced Planks constant, respectively.[37] This corresponds to a transmittance of $T_{ML} = 0.977$ which is equivalent to a monolayer absorbance value of $A_{ML} = -log(T_{ML}) = 0.0101$. We can convert this to an absorption coefficient by dividing by the thickness of a graphene monolayer, $t_{ML} = 0.35$ nm: $\alpha = A_{ML}/t_{ML} = 2.89 \times 10^7$ m$^{-1}$. This absorption coefficient should apply to a vertical stack of graphene sheets with centre-to-centre separation of 0.35 nm. We will use this known value of $\alpha$ for graphene to confirm the ability of our liquid-interface deposition technique to deposit a precise material thickness with each subsequent layer.

The UV-visible optical absorbance spectra for 1-6 graphene layers are displayed in Figure 3A. A peak at around 275 nm can be attributed to the $\pi-\pi^*$ electronic transition in graphene.[38] However, between 750 to 830 nm, the absorbance becomes almost independent of wavelength as expected for graphene.[39] Figure 3B shows a plot of absorbance taken at 800 nm (i.e. in the wavelength-independent regime), versus network thickness with the slope of this plot yielding the absorption coefficient, $\alpha = 1.51\pm0.09\times10^7$ m$^{-1}$. This value is lower than the value quoted above, a discrepancy which is most likely due to increased separation between the nanosheets caused by trapped solvent or hydrocarbons or indeed small pores within the film. We note that the conductivity of these graphene networks is likely limited by these trapped contaminants as they are expected to increase the resistance associated with inter-nanosheet transport. One can estimate the volume fraction of graphene within the film via the ratio of the measured to expected absorption coefficients to be 1.51/2.89=0.52, consistent with an enhanced mean separation between layers.

It is important to assess the ability of our deposition process to produce homogenous tiled monolayers over a large area. To achieve this, we performed high-resolution (nominally 10.6 µm), spatially resolved optical transmission measurements on our graphene networks using an optical transmission scanner. Typical images obtained by the scanner are included in the Supplementary Information (Figure S2), and a description of the procedure to manage the data is included in Methods. Briefly, we converted the transmission of each pixel to absorbance (A=-logT), subtract off the absorbance associated with the substrate, and then converted the pixel absorbance values to local network thickness ($t_{Net}=A/\alpha$). Although such data can provide information on the spatial variation of film thickness,[40] the simplest way to analyse it is to generate a histogram of all local network thickness values. Such histograms are shown in Figure 3C for 1L to 6L graphene networks. Each histogram is reasonably narrow, showing good thickness uniformity. We can quantify this via the ratio of thickness standard deviation to mean thickness ($\Delta t_{Net}/\langle t_{Net} \rangle$) which was 14% for 1L but typically 6-8% for the 2L-6L films (Figure 3C Inset). These low values of $\Delta t_{Net}/\langle t_{Net} \rangle$ show that our interfacial deposition methods yield uniform films. These histograms also display almost complete separation between peaks. Despite the ultrathin nature of each layer, this result shows that the deposition process produces homogeneous monolayer networks over a large area (square centimetre scale), and subsequent depositions add to the thickness of the network, without significant peeling or restacking in previous layers.

We measured the DC conductivity of our graphene networks, $\sigma_{Net}$, using a 4-point probe system to remove contact resistance (see Methods). Figure 3D shows the thickness-dependent conductivity of our networks. We find a sharp increase in network conductivity with increasing thickness from ~3400 S/m for the 1L network to ~$5\times10^4$ S/m for the 4L network, after which the conductivity saturates. Such thickness-dependent conductivity has been widely observed in nano-structured films and networks. In thin films of continuous materials (e.g. metal films), increasing and then saturating conductivity has been associated with surface scattering and roughness[41, 42] while in networks of discrete particles, such behavior is usually linked to disorder or percolation effects[43, 44]. While all of these effects may indeed be present here, our layer-by-layer assembled films probably display an additional characteristic which would lead to thickness dependent conductivity. It is well known that nanoparticle networks tend to be electrically limited by the presence of inter-particle junctions.[5, 28, 45, 46] In the case of a 1L network, we expect these junctions to be mostly nanosheet edge-edge junctions. However, for 2L and thicker networks, we also expect plane-plane junctions to play a role. Assuming plane-plane junctions have a lower junction resistance than edge-edge junctions, we expect an increase in network conductivity with increasing network thickness due to the growing dominance of low-resistance plane-plane junctions.

Our graphene networks have maximum measured network conductivity of $\sigma_{Net} = 5 \times 10^4$ S/m. This is among the highest reported in the literature, where reported values typically fall between $1.3 \times 10^3$ and $1.2 \times 10^5$ S/m.[27, 47-52] However, our networks are very unusual in that the conductivity reaches a maximum value at a thickness of 9 nm with 4 deposited layers. This is considerably thinner than the saturation thicknesses previously reported for graphene conductive networks in the literature, which are in the range 40 to 650 nm as summarised in Table S1.[27, 47-49, 53, 54].

The conductivity of a nanosheet network is currently limited by the inter-nanosheet junction resistance, $R_J$, with the relationship between these given by[28]

$$\sigma_{Net} \approx \frac{\sigma_{NS}(1-P_{Net})}{[1+2\sigma_{NS}t_{NS}R_J]} \quad (1)$$

Where $t_{NS}$ and $\sigma_{NS}$ are the nanosheet thickness and conductivity, respectively, and $P_{Net}$ is the network porosity. We can estimate $\sigma_{NS}$ to be the in-plane conductivity of graphite which has been reported in the range $\sim 10^5$-$10^6$ S/m.[55] However, from Raman spectroscopy (Figure S3), we know our nanosheets to be quite defective, so we can assume the conductivity is at the lower end: $\sigma_{NS} \sim 10^5$ S/m. Estimating the network porosity[27] at $P_{Net} \sim 0.25$ and using our maximum network conductivity of $\sigma_{Net} = 5 \times 10^4$ S/m, we estimate a minimum value of $R_J \approx 1920\ \Omega$. This is lower than the resistance of the component graphene nanosheets, which we calculate[28] to be $R_{NS} = (2\sigma_{NS} t_{NS})^{-1} = 3850\ \Omega$ (using $t_{NS}$=1.3 nm). This yields a value of $R_J / R_{NS} = 0.5$. We believe this is the first report of printed networks with $R_J < R_{NS}$. Achieving $R_J < R_{NS}$ is an important milestone in printed electronics since it facilitates networks with electronic properties approaching the individual nanosheets.[28]

*MoS$_2$ Stacked Monolayers – Optoelectronic Properties*

We now turn to the liquid interface deposited films of MoS$_2$, the characterisation of which was presented in figure 1B and E and figure 2C. We first measured the UV-visible absorbance spectra in MoS$_2$ films comprising 1 to 6 layers as shown in Figure 4A. These spectra show no significant shape changes but a general increase in absorbance with network thickness. The lack of any appreciable shift in excitonic peak position with layer number implies that there is minimal interlayer coupling in our networks.[56]

Analysis of Raman mapping spectra of our 1L MoS$_2$ network, taken on a 11 × 11 grid over a scan area of 1 mm$^2$ (Figure S4) shows a rather small separation between the $A_{1g}$ ($A_1$') and $E_{2g}$ ($E'$) peaks of between 15 and 17 cm$^{-1}$, This is slightly smaller than the separation of ~19 cm$^{-1}$ measured for monolayer MoS$_2$ exfoliated from bulk.[57] Furthermore, this 1L network shows high photoluminescence activity over the same 1 mm$^2$ area (Figure S4). A typical photoluminescence spectrum measured on a 1L network is shown in figure 4B. The 'main' peak centred at 1.82 eV (678 nm) and the 'shoulder' peak at 1.97 eV (630 nm) correspond to the characteristic A and B excitons of monolayer MoS$_2$, respectively.[58] . Since the PL intensity of single layer MoS$_2$ nanosheets is known to be much larger than that of bilayers and no indirect transition is observed here,[59, 60] this shows that the monolayers in the inks preserve their identity after deposition of the first layer, even over a large area. Only a few individual spots show few-layer behaviour, see Figure S4.

Plotting the absorbance at 2 eV, $A_{2eV}$, *vs* network thickness, $t_{Net}$, as shown in Figure 4C, showed a well-defined linear behaviour consistent with $A_{2eV} = \alpha_{2eV} t_{Net}$ where $\alpha_{2eV}$ is the absorption coefficient at 2 eV. From the slope of this graph, we found the absorption coefficient to be $\alpha_{2eV} = 1.55\pm0.13\times10^7$ m$^{-1}$. This shows good agreement with the literature values, which fall between $10^7$ to $10^8$ m$^{-1}$ in the optical wavelength range 1.5-3 eV. [61] [62] [63]

As with the graphene networks, the homogeneity of our MoS$_2$ networks were analysed using optical transmission scanner analysis. This procedure yields local thickness histograms consisting of well-defined sharp peaks (Figure 4D). While the peaks for the 1L to 4L networks are well separated, those peaks corresponding to the 4L and 5L networks show some broadening and overlap, suggesting increased roughness. However, we propose that such issues can be avoided with further process development. From the histograms, we calculate $\Delta t_{Net}/\langle t_{Net}\rangle$ for each film, finding values which fall from ~15% for the 1L film to ~6% for the 6L film (Figure 4D inset).

We performed a number of electrical measurements on our 1L to 6L MoS$_2$ networks in air and at room temperature. Due to the highly photosensitive nature of MoS$_2$ networks, and the presence of effects such as persistent photoconductivity,[64, 65] we performed electrical conductivity measurements under light soaking at power density of ~ 50 W.m$^{-2}$. Initial measurements were made by measuring the resistance of films while varying the channel length and these showed the presence of a non-trivial contact resistance, R$_C$ (figure 4E) (for further detail see Supplementary Information). Thus, to avoid the effects of contact resistance, we measured sheet resistance, $R_S$ using a 4-electrode methodology in Van Der Pauw (vdP) electrode configuration (electrode separation ~1 cm), which was shown to give equivalent values to the linear 4-point probe method (Figure S9). We note that the data in Figure 4C shows our films to be quite uniform when probed at a length scale of ~ 10 µm. This means they should be uniform enough to give good quality vdP measurements over a sample size of ~ 1 cm. The sheet resistance, $R_S$ was converted to DC conductivity of the network, $\sigma_{Net}$ via the network thickness $\sigma_{Net} = (R_S t_{Net})^{-1}$. The resultant conductivity is plotted as a function of network thickness in Figure 4F. As with our graphene networks, we find a considerable increase in network conductivity with thickness over the first three layers. We further find that the conductivity of the MoS$_2$ networks saturates for the 3-layer stacked monolayer network at an exceptionally high value of $\sigma_{Net} = 4 \times 10^3$ S/m

Previous work on aligned networks of electrochemically exfoliated MoS$_2$ showed much lower network conductivities of $\sigma_{Net} \approx 40$ S/m.[28] There are a number of reasons for this. It is likely that the light soaking resulted in elevated carrier densities, thus increasing the conductivity considerably. However, highly aligned nanosheets with large area junctions will result in low junction resistances, which will also result in high conductivity. We can use equation 1 to estimate $R_J$. To do this, we assume a carrier density of $10^{25}$ m$^{-3}$ (see below) and use the MoS$_2$ nanosheet mobility as measured by THz spectroscopy[28] (40 cm$^2$/Vs) to estimate the conductivity of individual nanosheets to be $\sigma_{NS} \approx 6.4 \times 10^3$ S/m. Again, this value is high, due to the high carrier density. We note that our measured network conductivity of $4 \times 10^3$ S/m is close to this value, consistent with our MoS$_2$ network having very low junction resistances.[28] Then, using these values in equation 1, and assuming network porosity[27] of $P_{Net} = 0.25$ gives a junction resistance of $R_J \approx 26$ kΩ. This value is much lower than the value of $R_J \approx 2$ MΩ, previously reported for networks of electrochemically exfoliated MoS$_2$.[28] Again, this value is lower than the resistance of the individual MoS$_2$ nanosheets, which we calculate[28] to be $R_{NS} = (2\sigma_{NS} t_{NS})^{-1} = 130$ kΩ (using $t_{NS}$=0.6 nm). This data implies $R_J / R_{NS} \sim 0.2$. We reiterate that achieving $R_J < R_{NS}$ is particularly important for printed semiconductors as it allows the maximisation of network mobility.[28]

As indicated above, this very low junction resistance is at least partly due our large area junctions. However, we also note that previous work shows a well-defined, near linear scaling of $R_J$ with $\sigma_{NS}^{-1}$.[28] Thus, the increased values of $\sigma_{NS}$ due to light soaking should result in a reduction in $R_J$ alongside that associated with junction morphology.

The field-effect mobility of our MoS$_2$ networks was determined by fabricating thin film transistors (TFTs) with the MoS$_2$ stacked monolayer networks as the channel material, as outlined in Methods. Measurements were made for various channel lengths. We performed field-effect measurements by applying gating both dielectrically via a 230 nm thick SiO$_2$ layer and electrochemically via an ionic liquid. Typical measurements for a 2L network are shown in Figure 5A (see Figure S8 for more transfer curves). This graph shows the electrochemically gated network to show n-type behaviour with an on-off current ratio of >2000. However, for the dielectrically gated network, the source-drain current changes only weakly with gate voltage. The reason for this difference is that the oxide in the dielectrically gated sample is relatively thick (~230 nm), leading to an areal capacitance that is low ($1.38 \times 10^{-4}$ F/m$^2$) compared to that associated with the electrochemical double layer in the electrochemically gated network ($2.48 \times 10^{-2}$ F/m$^2$).

We can illustrate this as follows. The source drain current in the linear region is given by

$$I_{ds} = \frac{\mu_{Net} V_{ds} W C_A}{L} \left( V_g - V_T - V_{ds}/2 \right) \qquad (2)$$

where $\mu_{Net}$ is the network mobility, $L$ and $W$ are the channel length and width, respectively, $V_{ds}$ is the applied drain-source voltage, $V_T$ is the threshold voltage and $C_A$ is the areal capacitance of the dielectric (or double layer in the case of electrochemical switching). This means that, in the linear region, $I_{ds}$ should scale linearly with $C_A V_g$. Thus, we can correct for the difference in capacitances (essentially normalising with respect to dielectric thickness) by plotting $I_{ds}$ versus $C_A V_g$ as shown in figure 5B. Here we see that the electrochemically gated transfer curve follows the expected linear form. In addition, the dielectrically gated data consists of a linear curve close to, and with slope similar to, the electrochemically gated transfer curve. This shows the dielectrically gated data to be in the linear regime and confirms the differences between the curves to be predominately due to differences between the dielectric and the electrochemical double layer. Analysis of these curves showed the network mobilities to be very similar: 9 cm$^2$/Vs for the dielectrically gated network and 7 cm$^2$/Vs for the electrochemically gated network.

We subsequently measured the mobility for all samples via dielectric gating. The apparent mobility was calculated from the TFT transfer curves using the formula:

$$\mu_{Net} = \frac{dI_{ds}}{dV_g} \frac{L}{V_{ds} W C_A} \qquad (3)$$

Where $dI_{ds}/dV_g$ is the slope of the transfer curve in the linear region (i.e. the transconductance), We use the term apparent mobility as the value in this way is artificially reduced due to the presence of contact resistance. The resultant mobility is plotted versus network thickness in Figure 5C (open symbols). As with the conductivity, the apparent mobility increases with network thickness, reaching a maximum value of around 11 cm$^2$/Vs for the thickest films. This value is on among the highest reported in literature for solution processed MoS$_2$ TFTs, which fall between 0.01 and 11 cm$^2$/Vs (A table of literature reported MoS$_2$ TFTs is provided in Supplementary Information Table S2).[20, 22, 30, 32, 66, 67] However, this is somewhat lower than the aforementioned value of $\mu_{NS}$=40 cm$^2$/Vs measured by terahertz spectroscopy.[28] Given that there is a considerable contact resistance in these films and we know that the junction resistance is very low, it is likely that the true network mobility is actually considerably higher than this value.

As mentioned previously, our MoS$_2$ TFTs demonstrate significant contact resistance, $R_C$. To eliminate these effects from our TFTs, we used relatively simple mathematical approaches to correct the measured mobility values. One method involves fitting the channel length dependence of the measured mobility to a simple model based on the transfer length method (TLM). The other method involves correcting the applied voltage for the voltage drop across contacts. Both are described in the Supplementary Information. The resultant average $R_C$-corrected mobility is plotted in Figure 5C as a function of network thickness. We find the $R_C$-corrected mobility to increase with network thickness, reaching 30 cm$^2$/Vs for the 4L and 5L networks. It is important to note that this value is quite close to the value measured for individual nanosheets quoted above ($\mu_{NS}$ = 40 cm$^2$/Vs). This is entirely consistent with the presence of low junction resistance and the fact that as $R_J$ gets small relative to the nanosheet resistance, $R_{NS}$, the network mobility should approach that of the individual nanosheets (expressed via $\mu_{NS}/\mu_{Net} \approx R_J/R_{NS} +1$)[28]. Taking the values of $R_{NS}$ and $R_J$ estimated above gives $R_J/R_{NS} \sim 0.2$, implying that $\mu_{Net}$ should be only slightly smaller than $\mu_{NS}$: $\mu_{Net} \approx \mu_{NS}/1.2$. Assuming[28] $\mu_{NS}$ = 40 cm$^2$/Vs then gives $\mu_{Net} \sim$ 33 cm$^2$/Vs consistent with the value measured for the for the 4L and 5L networks.

It is also worth noting that our measured network mobilities, $\mu_{Net} \sim$30 cm$^2$/Vs, are, to our knowledge, the highest reported values for solution-processed MoS$_2$ networks which usually display mobilities in the range of 0.1 to 11 cm$^2$/Vs as previously mentioned. In addition, our network mobility is comparable with some values reported for TFTs made from individual MoS$_2$ flakes, whereby single-crystal MoS$_2$ transistors without passivation show mobilities 0.1-10 cm$^2$/Vs,[68, 69] while mobilities of around 27-28 cm$^2$/Vs were achieved when care was taken to avoid short-channel effects,[70] or substrate effects.[71]

We can estimate the carrier concentration in our MoS$_2$ networks by combining the ($R_c$-corrected) conductivity data in figure 4D with the $R_c$-corrected mobility data in figure 5C. As shown in Figure 5D, $n$ shows a dramatic increase from 3 $\times 10^{23}$ $m^{-3}$ for the 1L to $\sim 10^{25}$ $m^{-3}$ for the thicker networks. The values for thicker networks are considerably higher than previously reported values for networks of similar nanosheets ($\sim$5 $\times 10^{23}$ $m^{-3}$)[28]. We propose that this dramatic increase from 1L to 2L is linked to interactions with the substrate.[71] Finally, the overall carrier concentration for 2 to 5 layer MoS$_2$ networks is very high even compared to literature examples,[72] which tend to show high levels of doping in the MoS$_2$ networks. These high carrier densities are probably associated with a considerable doping effect caused by the IPA treatment of our MoS$_2$ networks during fabrication. This kind of doping acts to reduce the on-off ratio in TFTs. However, we note that the MoS$_2$ networks can be effectively de-doped using a previously reported strategy of soaking of the network in acetone for 20 minutes followed by chloroform for 20 minutes.[73] In this way, the on-off ratio of our $MoS_2$ transistors can be improved by approximately 4 orders of magnitude, and the doping concentration can be reduced by almost an order of magnitude (Figure S5).

*Quasi-2D Materials: Ag nanosheets optoelectrical properties*

To emphasise the utility of our liquid-interface deposition method beyond standard 2D materials, we now turn to films of the non-layered, quasi-2D colloidal silver nanosheets previously described in figure 1C and F. The UV-visible reflectance spectra measured for Ag nanosheet networks with thickness 1L to 3L, before annealing at 250 °C, is given in Figure 6A. It can be seen by eye that these networks of AgNS deposited on glass are highly reflective; even a monolayer of AgNS demonstrates a high degree of specular reflection as shown in Figure 6B. Although the 1L film shows considerable reflectance, the 2L and 3L films are even more reflective, both showing reflectance of 70-80% over the visible spectrum. The origin of the optical differences between the 1L and the 2L and 3L films can be attributed to the ~10 % of additional light transmitted through in the 1L film (SI Figure S6). This corresponds well with the difference in uncovered area from 1L to 2L AgNS network, as determined by statistical analysis of SEM micrographs (SI Figure S6).

To determine the conductivity of the 1L, 2L and 3L films of AgNS, we first anneal the networks under inert atmosphere at 250 °C to remove excess surfactant from the surface of nanosheets and sinter the networks[35] (Plots of conductivity versus annealing temperature are provided in Figure S7). We then use 4-probe techniques to determine the sheet resistance of the annealed networks. As shown in figure 6C, the conductivity of annealed 1L network was low, $\sigma_{Net}$ ~$10^{-3}$ S/m. However, going from 1 to 2 layers, the network conductivity increases by over 9 orders of magnitude.

Again, SEM images of 250 °C annealed 1L and 2L films suggest the cause of this (Figure 6D and E, respectively). While the as-deposited Ag monolayer networks show good edge-edge connectivity over a large area, these images show that annealing causes the AgNS to partially melt and 'ball up' due to surface tension effects, resulting in a disconnected array of Ag islands (Figure 6C). In contrast, the bilayer network remains connected after annealing, possibly due to out-of-plane sintering and improved thermal conduction in the bilayer network. This dramatic change in morphology between the 1L and 2L annealed networks is responsible for the significant increase in conductivity.

There is only a small variation in electrical conductivity between our 2L and 3L networks, with the 3L film displaying $\sigma_{Net} = 7\pm3\times10^7$ S/m, close to the level of bulk silver. This is among the highest values reported for printed silver networks. For example, Lee et. al. [74] found that the conductivity of AgNS network, $\sigma_{Net} = 1.4 \times10^7$ S/m was greater than that of 0D silver nanoparticles (AgNP) networks, $\sigma_{Net} = 3.3 \times10^6$ S/m owing to the larger area nanosheet junctions in the AgNS network.[74] Conversely, Stewart et. al. found the opposite trend,[75] with their drop-cast AgNS networks reaching maximum conductivity of $\sigma_{Net} = 2.4 \times10^5$ S/m, less than AgNP networks deposited and processed in the same way, $\sigma_{Net} = 5 \times10^6$ S/m. [75] Tai et. al. reported AgNS networks deposited by a modified gel pen, which displayed network conductivity of $\sigma_{Net} = 1.1 \times10^7$ S/m. [76] Finally, Kelly et. al. displayed thickness-dependent conductivity in aerosol-jet printed AgNS networks, which reached thickness-independent maximum conductivity of $\sigma_{Net} = 1.3 \times10^7$ S/m at thickness $t_{Net} = 140$ nm. [35] That we measure network conductivity $\sigma_{Net} = 3.2 \times10^7$ at a thickness of ~90 nm is testament to the ideal morphology, in particular a reduced porosity and improved interlayer interfaces, achievable through liquid interface deposition.

CONCLUSION

In 2D materials networks, the ratio of inter-nanosheet junction resistance to nanosheet resistance, $R_J/R_{NS}$ is a predictor of network electrical performance. Ultimately, before the integration of 2D materials in many printed electronics applications, strategies to achieve $R_J/R_{NS} \ll 1$ must be realised. In this work, we have utilised a liquid interface assembly-based deposition process to deposit nanosheet networks of graphene, $MoS_2$ and AgNS with maximised nanosheet alignment and uniformity. Thus, allowing us exceptional control of nanosheet junction morphology. We confirm the ideal morphology in these networks, systematically measure their thickness-dependant electrical performance and, for the first time, report $R_J/R_{NS} < 1$. This allowed us to achieve record-high mobility in $MoS_2$ networks (contact resistance-corrected), and among the highest reported conductivity in graphene and AgNS networks but at record-low network thickness.

This work highlights the importance of morphology in maximising the electrical performance of nanosheet networks and serves as a guide to future works on 2D materials-based printed electronics. Further improvements to nanosheet network performance may be achievable by controlling morphology in conjunction with chemical crosslinking or doping strategies, and we are currently exploring these avenues.

METHODS

*Synthesis of 2D materials dispersions:*

To prepare graphene nanosheet dispersions, two pieces of graphite foil (alfa aesar), 50×30×0.25 mm$^3$ were clamped to crocodile clips connected to a DC power supply as cathode and anode and immersed in 100 mL of 0.1 M aqueous $(NH_4)_2SO_4$, separated by a fixed distance of 2 cm. A potential of 10 V was applied for 30 min, with the current rising from ~1 A to ~2 A during the process. The resulting material (expanded graphite from the anode) was filtered and washed with ~1 L of Milli-Q water to remove residual electrolyte, and then bath sonicated in 100 mL DMF for 10 min, forming a stable ink.

$MoS_2$ nanosheet dispersions were prepared using a previously-reported electrochemical intercalation procedure.[22] Briefly, a 2-electrode electrochemical cell was set up in a 50 mL beaker. A strip of graphite foil (1 mm thick, 97 % (metal basis), Thermo Scientific) was used as the anode and a 1 mm thick, 0.5 cm$^2$ $MoS_2$ single crystal (natural origin, Krupka, Czech Republic) was used as the cathode. The electrodes were submerged in an electrolyte consisting of tetraheptylammonium bromide $[THA]^+ [Br]^-$ (Sigma-Aldrich) in acetonitrile (≥ 99.5 %, Sigma-Aldrich) at a concentration of 12.5 mg/mL. A potential of 7 V was applied across the electrodes for 1 h. During this time, the $MoS_2$ expands due to intercalation with $[THA]^+$. The intercalated $MoS_2$ crystal was rinsed with acetone to remove $[Br]^-$ and then sonicated in a 2 % w/v solution of polyvinyl pyrrolidone (PVP) (40,000 g/mol, Sigma-Aldrich) in dimethylformamide (DMF) (≥ 99 %, Sigma-Aldrich) for 30 minutes to obtain a dark green dispersion.

Ag nanosheet inks were prepared by diluting commercially-sourced stock aqueous dispersion (N300 Nanoflake, Tokusen, USA) to a concentration of 80 – 100 mg/mL with DI water (18.2 MΩ.cm).

*Size selection of 2D materials dispersions:*

The distribution of nanosheet thicknesses in each 2D material dispersion was narrowed by cascade centrifugation[77] (Hettich Mikro 220R, fixed-angle rotor, Massachusetts, USA). For each dispersion, the large and thick nanosheets were sedimented at [force 1] for [time 1] and the sediment was discarded. The supernatant was then decanted and subjected to a further centrifugation step at [force 2] for [time 2]. The small and thin nanosheets in the supernatant were discarded and the sediment was redispersed in in neat IPA (graphene and $MoS_2$) or 2:1 volume ratio mixture of IPA:DI water (Ag nanosheets) to produce the size-selected ink.

Ag nanosheets: [Force 1: 112 g Time 1: 2 h], [Force 2: 447 g Time 2: 2 h]

MoS$_2$: [Force 1: 958 g Time 1: 30 min], [Force 2: 3830 g Time 2: 60 min]

Graphene: [Force 1: 958 g Time 1: 20 min], [Force 2: 3830 g Time 2: 90 min]

*Deposition of liquid-interface assembled monolayers:*

Deionised water (40 mL, >18 MΩ.cm) and n-hexane (10 mL, ≥ 99 %, Sigma-Aldrich) were added to a 50 mL beaker with magnetic stirrer bar set stirring at approximately 100 RPM (Micro Stirrer F203A0440, Kleinfield, Gehrden, Germany). A suitable was placed on a custom substrate stage, fabricated from 1mm PTFE sheet (Bohlender GmbH, Gruensfeld, Germany). The substrate stage was mounted in a dip coater (Dip Coater, Ossilla, Leiden, NL) and submerged beneath the hexane-water interface. 2D materials dispersions in IPA or IPA/H$_2$O mixtures were injected at the interface of hexane and water at a rate of 150 μL/min using a syringe pump (SPM, DK Infusetek Co, Shanghai). Stirring was stopped 5 minutes after the assembly of a complete monolayer at the interface, and deposition of the monolayer was achieved by lifting the substrate through the interface at a rate of 1 mm/s.

*Microscopic characterisation:*

Atomic force microscopy was carried out using a Multimode 8 Atomic Force Microscope (Bruker, MA, USA). Scanasyst mode and a Bruker Scanasyt cantilever was used for all images. For the AFM measurement of individual nanosheets, 2D materials dispersions were drop cast onto (3-Aminopropyl)triethoxysilane (APTES)-coated silicon wafer substrates (2000 nm thermally-grown oxide, SK siltron Ltd., Gumi, Korea). NanosheFet heights were measured by drawing line profiles from bare substrate to centre of nanosheet and fitting using the 'critical dimension' feature of Gwyddion software. Nanosheet lengths and widths were obtained from the same AFM images by measurement using the FIJI software.[78] Nanosheet length was defined as the longest line that can be drawn across a given nanosheet. Nanosheet width was defined as the longest line that can be drawn perpendicular to the length. AFM measurements of nanosheet height give an apparent thickness rather than the real nanosheet thickness,[79-82] possibly due to tip-sample contrast issues or trapped solvent. In particular, the SI of ref[83] contains a good summary of this topic. To convert the measured apparent thickness to the real nanosheet thickness, one divides the measured thickness by the apparent thickness per monolayer (ML) which has been measured for a range of 2D materials.[83] This quantity is 0.95 nm/ML and 1.9 nm/ML for graphene and MoS$_2$ respectively. This yields the number of ML per nanosheet. This parameter can then be multiplied by the real thickness per ML (0.35 and 0.7 for graphene and MoS$_2$ respectively) to get the real nanosheet thickness.

Scanning electron microscopy was carried out using a Zeiss Ultra FEG SEM. Samples were not coated prior to imaging. An accelerating voltage of between 1 to 5 keV, and a working distance of between 3 and 5 mm was used in all images.

Cross-sectional FIB lamellae were fabricated using an FEI Helios NanoLab 660 dual beam FIB, and TEM images obtained with an FEI Talos F200A TEM at 200 kV. The samples were coated with approx. 20 nm of carbon prior to FIB processing to reduce charging and protect the $MoS_2$ layer. Target area was coated with 3 μm of Pt via electron beam and ion beam deposition, before a bulk mill at 30 kV and sample thinning at voltages down to 5 kV to achieve electron transparency.

*Optical Characterisation:*

UV-visible absorbance spectra were obtained using a Lambda 1050 Spectrophotometer (Perkin Elmer, MA, USA). Wavelength range of 250 to 830 nm at wavelength step size of 0.5 nm. Extinction, scattering, and reflectance spectra were obtained for each film and absorbance was calculated by subtracting the scattering and reflectance components from extinction.

Raman and PL measurements of $MoS_2$ films were performed using a LabRam spectrometer (HORIBA Jobin Yvon GmbH) in backscattering geometry and a laser wavelength of 532 nm in ambient conditions. The emitted Raman signal was collected through a 100× magnification objective and dispersed by a 1800 l/mm grating. The laser power was kept below 0.3 mW for all measurements to avoid local heating. All Raman spectra were calibrated using Neon lines. PL spectra were acquired in the same setup but with a 300 l/mm grating and laser power below 0.1 mW. The maps consist of 11×11 single-point measurements over an area of 1 mm² with a distance of 100 μm in between individual spots. Raman measurements of 1L graphene films were obtained using an Alpha300 R Raman microscope (WITec, GmbH, Germany). A 532 nm laser was used at power of < 1mW. The Raman signal was collected through a 10× objective and dispersed through a 1800 l/mm grating.

Optical transmission scanner images were obtained by mounting graphene or $MoS_2$-coated 2.5×2.5 cm glass substrates (Microscope Slides, Fischer Scientific) onto a drop of DI water (>18 MΩ.cm) on the bed of a Perfection V850 Pro scanner (Epson Corp., Suwa, Japan) at a resolution of 2400 dpi (~10.6 μm per pixel). The resulting images were saved as Multi-TIFF format and the blue channel was used for further processing. The blue channel pixel intensity, <px> was converted to transmission T using a quadratic relationship obtained empirically in our group via UV-visible spectroscopy of similar 2D materials networks: $T = -0.01622 + (4.1922 \times 10^{-6})<px> + (1.6598 \times 10^{-10})<px>$. Average transmission could then be converted to extinction values via the formula Ext = logT. The extinction of bare substrate was subtracted to obtain the film extinction, $Ext_{Film} = Ext_{Sample} - Ext_{Substrate}$. After, the peak positions were adjusted according to the extinction of the networks obtained via UV-visible spectroscopy. The extinction values could be converted to thickness using the extinction coefficient obtained via thickness-dependent UV-visible spectroscopy as shown in this work. A Python script was written to select at random 100000 extinction (thickness) values to reduce file size and further processing time. These datapoints were used to plot the histograms of network thickness shown in this work. Since this technique requires optically transmissive networks, Ag nanosheet networks were not analysed in this way.

*Optical Profilometry*

We obtained film thickness measurements using a 50× objective lens on a Profilm3D Optical Profiler (Filmetrics, KLA Corporation, USA). Graphene and $MoS_2$ networks on glass slides, were scratched with fine-tipped tweezers to expose the underlying substrate as a reference surface. White light interferometry mode was used to generate interference patterns for thickness determination. The 3D images generated were processed using the ProfilmOnline software. Each pixel height value was used to plot a histogram, which resulted in a bimodal distribution. The peak corresponding to substrate was subtracted from the peak representing the network to obtain the mean thickness of the network.

*Electrical Characterisation:*

To measure conductivity of graphene and Ag nanosheet networks, we used a commercial 4-point probe head (Four-Point Probe, Ossila, Leiden, NL). This system utilises four spring-loaded probes with rounded tips with 1mm probe spacing and applies a constant force whilst in contact. A Current was sourced across the outer electrodes and the voltage was measured between the inner two electrodes. The measured resistance, $R = V/I$ was averaged over 5 measurements at random points across each network. Resistance was converted to sheet resistance in line with the derivation by Smits.[84] $R \frac{\pi}{\ln 2} = R_s$

The sheet resistance of $MoS_2$ networks was measured by depositing Ag paste contacts onto the four corners (Van der Pauw configuration) of 2×2 cm square-shaped networks. The Ag paste was dried on a hot plate at 50 °C for 1h. The $MoS_2$ networks on glass substrates were subject to light soaking conditions (4320 lux, 48 W/m², RS PRO Light Box, RS Components, Northamptonshire, UK), and current was forced through two adjacent probes, while voltage was measured across the opposite probes. The device was rotated 90 and the measurement repeated. These perpendicular measurements generally differed by <30%. The averaged resistance was converted to sheet resistance using[84] $R_S = [(R_\parallel + R_\perp)/2] \times \pi / \ln 2$. We note that similar sheet resistance values for graphene were measured using both linear 4-point probe and Van der Pauw electrode arrangements (Figure S9). The sheet resistance of the graphene, MoS$_2$, and Ag nanosheets networks were converted to conductivity, $\sigma$ by considering the thickness of the networks, $\sigma = 1/R_S t$. The thickness of MoS$_2$ networks were averaged over 3 (graphene) or 4 (MoS$_2$) measurement methods; AFM, absorbance spectroscopy, optical profilometry and TEM cross-section imaging (MoS$_2$ only).

*MoS$_2$ TFTs:*

Back-gate and bottom-contact, dielectric gated MoS$_2$ TFTs were fabricated by depositing MoS$_2$ liquid-interface deposited monolayers (1L, 2L, 3L, 4L, 5L) directly onto n-doped silicon substrates with 230 nm SiO$_2$ gate oxide and ITO/Au source-drain contacts (Gen 5 OFET Test Chips, IPMS Fraunhofer, Dresden, Germany). These chips have prepatterned 4-contact electrode arrays, which have fixed channel width of 2 mm and channel lengths of 2.5, 5, 10 and 20 µm (four electrode arrays per channel length, see Figure S11). The devices were soaked in IPA at 80 °C for 30 minutes prior to electronic testing. We first measured the network resistance (averaged over four devices per channel length) for each channel lengths to extract the contact resistance. We then measured transfer curves (I$_d$ vs. V$_g$) using the back gate which is built into the test chips. For all five different network thicknesses (1L, 2L, 3L, 4L, 5L), this was carried out for all four channel lengths (four devices per channel length). Contact resistance correction was performed as described in the SI. After dielectric gating, we performed electrochemical gating experiments on a subset of the thinner devices. For these electrochemically-gated TFTs, a drop of 1-ethyl-3-methylimidazolium bis(trifluoromethylsulfonyl)imide (EMIM TFSI) (Sigma Aldrich) was applied to the surface of equivalent MoS$_2$ devices between the source and drain electrodes. A tungsten probe tip was submerged into the electrolyte droplet to act as the gate electrode. The EMIM TFSI electrolyte was dried under vacuum at 60 °C prior to use. Electronic testing of all MoS$_2$ TFTs was performed using a Janis ST-500 Probe Station (Lake Shore Cryotronics Ltd., OH, USA) in conjunction with a Keithley 4200A-SCS Parameter Analyser (Keithley Instruments. Ohio, US).

ACKNOWLEDGEMENTS: We gratefully acknowledge support from Horizon Europe (e.g. project 2D-PRINTABLE). We have also received support from the Science Foundation Ireland (SFI) funded centre AMBER (SFI/12/RC/2278_P2) and availed of the facilities of the SFI-funded AML and ARL labs. JN acknowledges funding from ADMIRE Marie Skłodowska-Curie COFUND Postdoctoral Fellowship 'SLAM-2D' (grant number 12/RC/2278_P2). T.C. acknowledge funding from a Marie Skłodowska-Curie Individual Fellowship "MOVE" (grant number: 101030735, project number: 211395, and award number: 16883). S.J.H. acknowledges from the engineering and physical sciences research council (EP/V007033/1, EP/V001914/1, EP/S021531/1, EP/M010619/1, and EP/P009050/1) and from the European Union's Horizon 2020 research and innovation programme (Grant ERC-2016-STG-EvoluTEM-715502). EM access was supported by the Henry Royce Institute for Advanced Materials, funded through EPSRC grants EP/R00661X/1, EP/S019367/1, EP/P025021/1 and EP/P025498/1. ES and JM acknowledge support from the Dr. Isolde Dietrich-Stiftung and the Deutsche Forschungsgemeinschaft (DFG, German Research Foundation), project number 447264071 (INST 90/1183-1 FUGG).

Figures

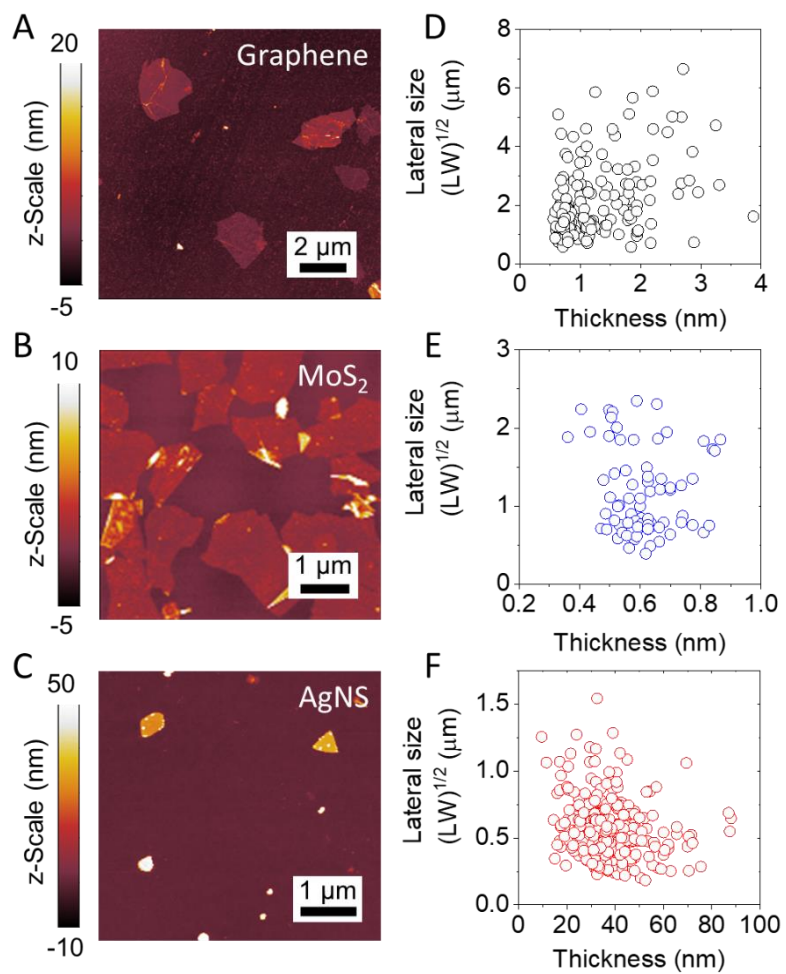

Fig 1. Size characterisation of 2D materials inks: A-C. AFM image of individual nanosheets in A. the graphene ink B. the MoS$_2$ ink and C. the AgNS ink. D-E. Scatter plots of lateral nanosheet length and thickness for individually measured nanosheets in D. the graphene ink [counts = 160] E. the MoS$_2$ ink [counts = 70] F. the Ag ink [counts = 300].

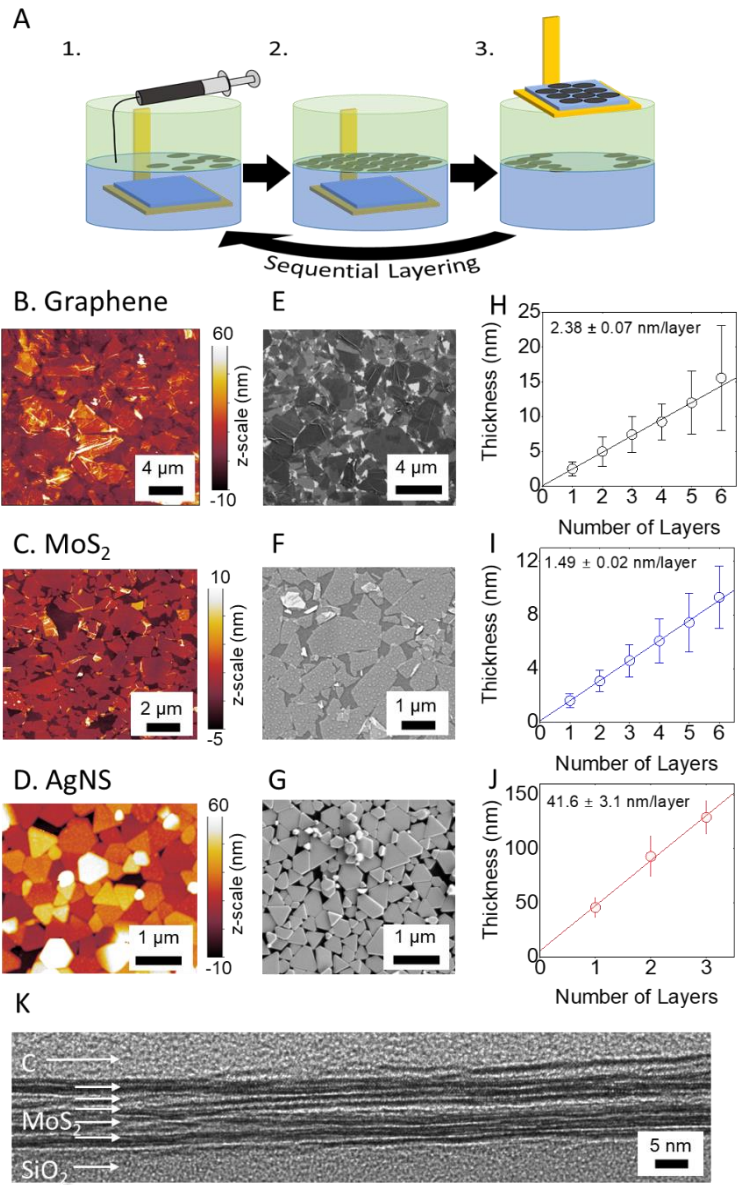

Fig 2. Liquid-liquid interface deposition: A. Schematic representation of the liquid-liquid interface deposition process: 1. Nanosheet ink is injected at the water-hexane interface until: 2. A complete monolayer is formed after which: 3. A substrate to be coated is lifted vertically through the interface to deposit the monolayer. The deposition process can be repeated for thicker networks. AFM micrographs of monolayer networks of B. graphene C. $MoS_2$ D. AgNS. SEM micrographs of monolayer networks of E. graphene F. $MoS_2$ and G. AgNS. Plots of monolayer thickness as a function of number of depositions for H. graphene I. $MoS_2$ and J. AgNS The red lines in H, I and J are linear fits to the data and the thickness per layer is indicated in the plots. The thickness in D and G were obtained by averaging values obtained via AFM and three other measurement techniques, as outlined in Methods. K. cross sectional TEM micrograph of a 5-layer $MoS_2$ network.

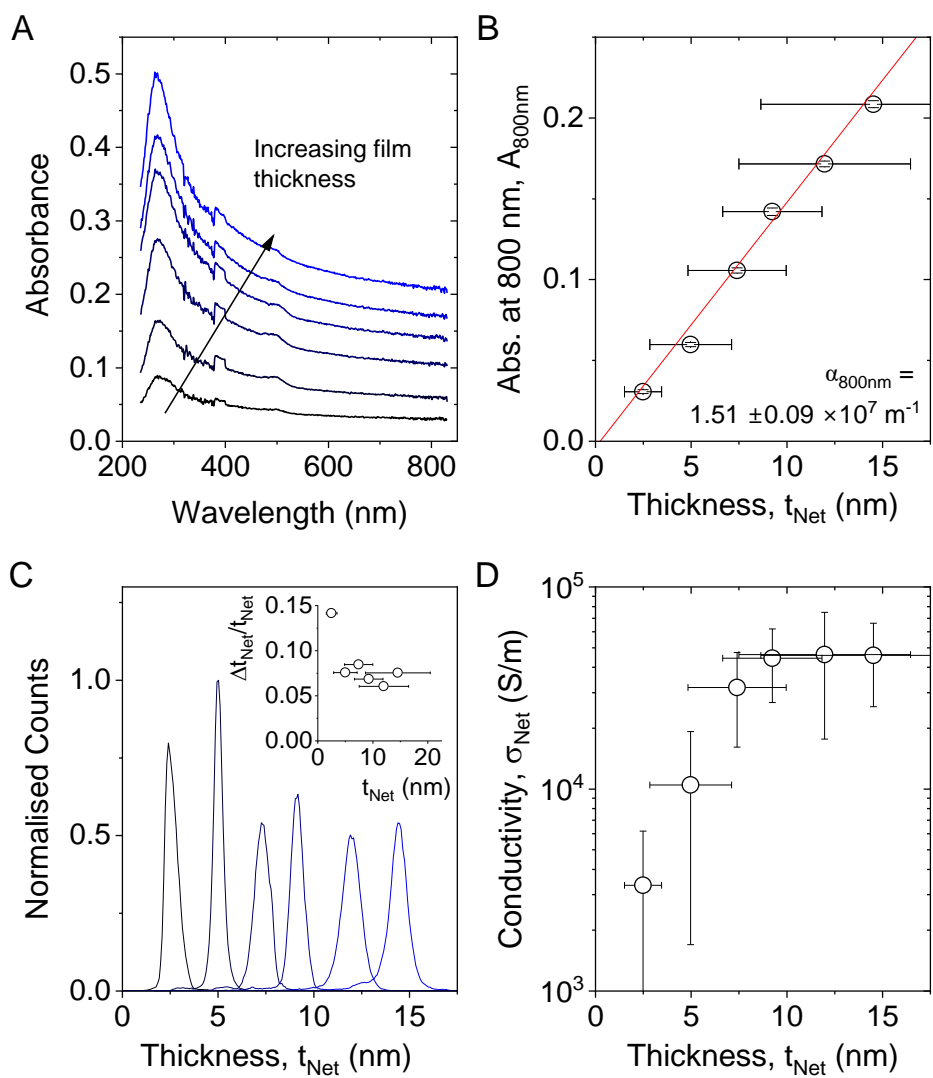

Fig 3. Optoelectrical characterisation of liquid-interface assembled graphene networks A. UV-visible absorbance spectra for graphene networks 1 to 6 layers thick. B. A plot of absorbance (at 800 nm) versus thickness. The extinction coefficient, $\alpha_{800nm}$ is indicated on the plot. C. Transmission scanner histograms showing local network thickness for Graphene networks from 1L to 6L. Inset: Ratio of standard deviation of network thickness to mean network thickness plotted versus mean network thickness. D. Plot of conductivity versus thickness for 1 to 6 layers of graphene.

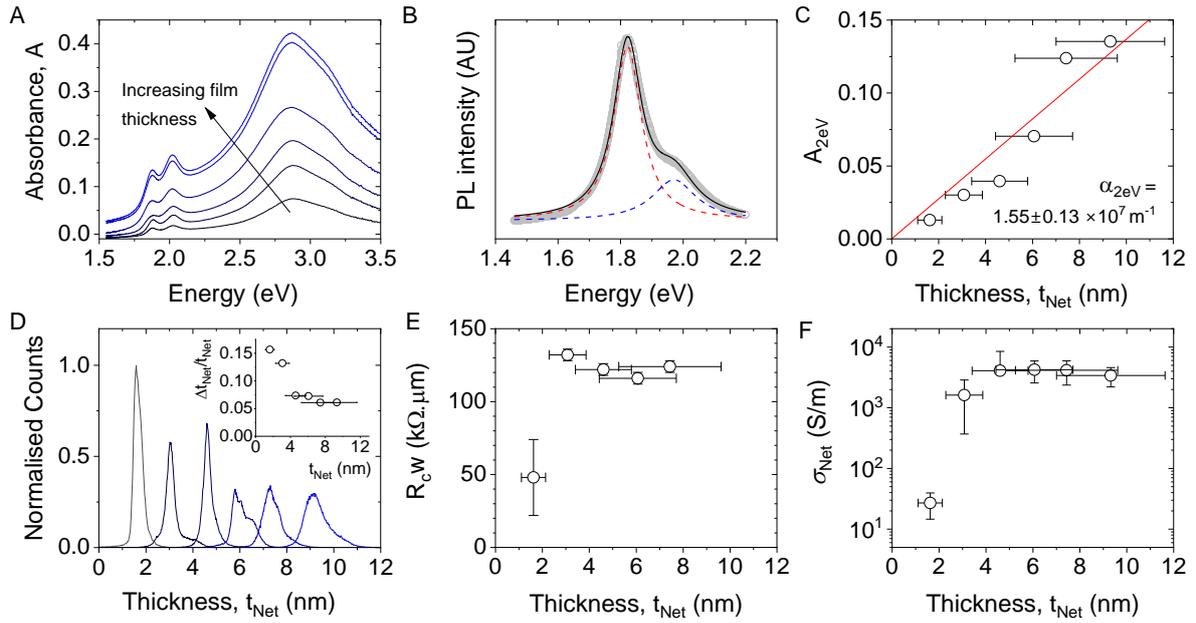

Fig 4. Optoelectronic properties of MoS$_2$ stacked monolayer networks. A. UV-visible absorbance spectra for MoS$_2$ networks 1 to 6 layers thick. B. The PL signal from a 1L MoS$_2$ film (gray circles). This spectrum has been fitted to two Lorentzian components (dashed lines). The overall fit is shown by the solid black line. C. Plot of absorbance measured at a photon energy of 2eV vs thickness for MoS$_2$ films. The red line is a linear fit to the data and the absorbance coefficient at 2eV, $\alpha_{2eV}$ is indicated in the plot. D. Transmission scanner histograms showing local network thickness for MoS$_2$ networks from 1L to 6L. Inset: Ratio of standard deviation of network thickness to mean network thickness plotted versus mean network thickness. E. Contact resistance, expressed as $R_C \times w$, where w is the channel width (2mm), plotted vs thickness. F. Conductivity versus thickness for 1L to 6L MoS$_2$ networks.

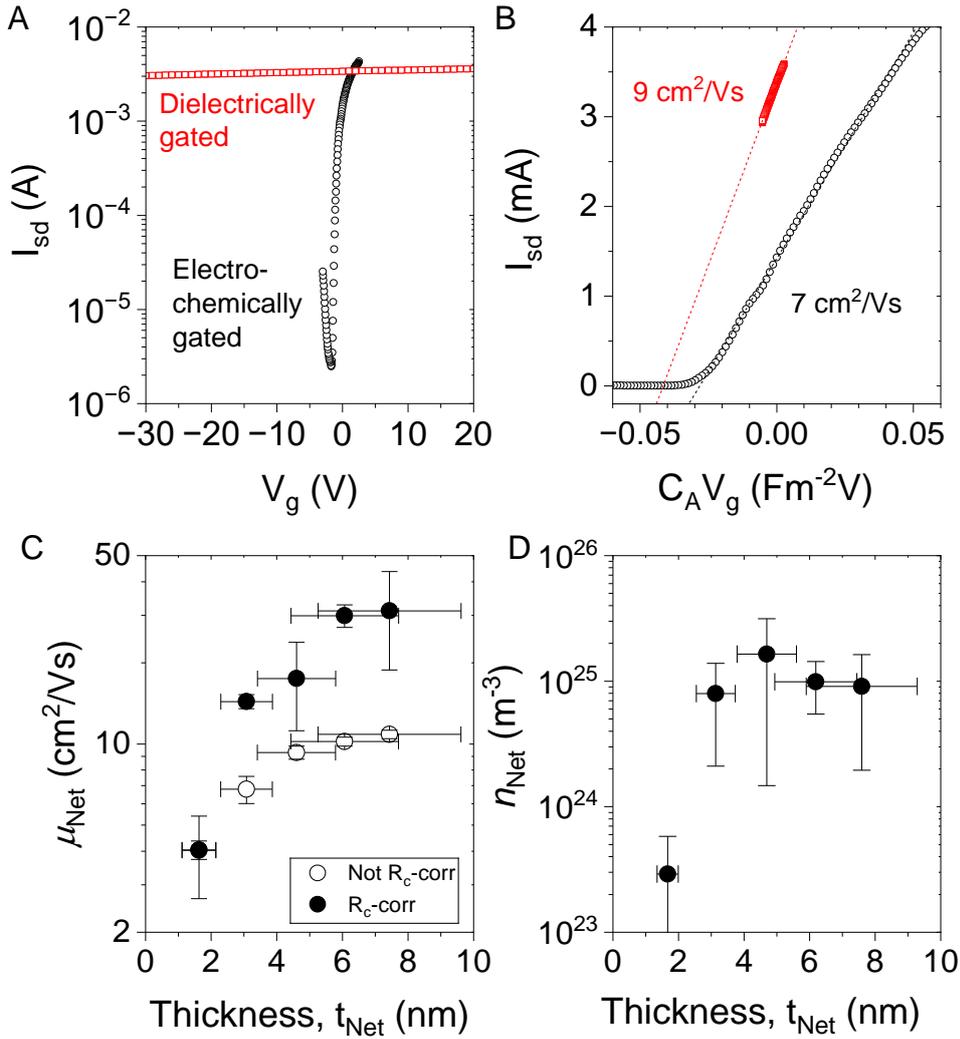

Fig 5. Mobility measurements in MoS$_2$ networks. A. Transfer curves for 2L MoS$_2$ networks (thickness 3.1 nm, channel length, L=20 µm and width, W=2 mm) in a field-effect transistor configuration gated dielectrically via a 230 nm thick SiO$_2$ layer (red) and electrochemically via an ionic liquid (black). B. Source-drain current plotted versus the product of areal capacitance and gate voltage for both samples shown in A. C. Average network mobility plotted as a function of network thickness as measured (open symbols) and after correction for the effects of contact resistance (solid symbols). D. Carrier density calculated from R$_c$-corrected mobility (Fig 4C) and R$_c$-corrected conductivity (Fig 3E).

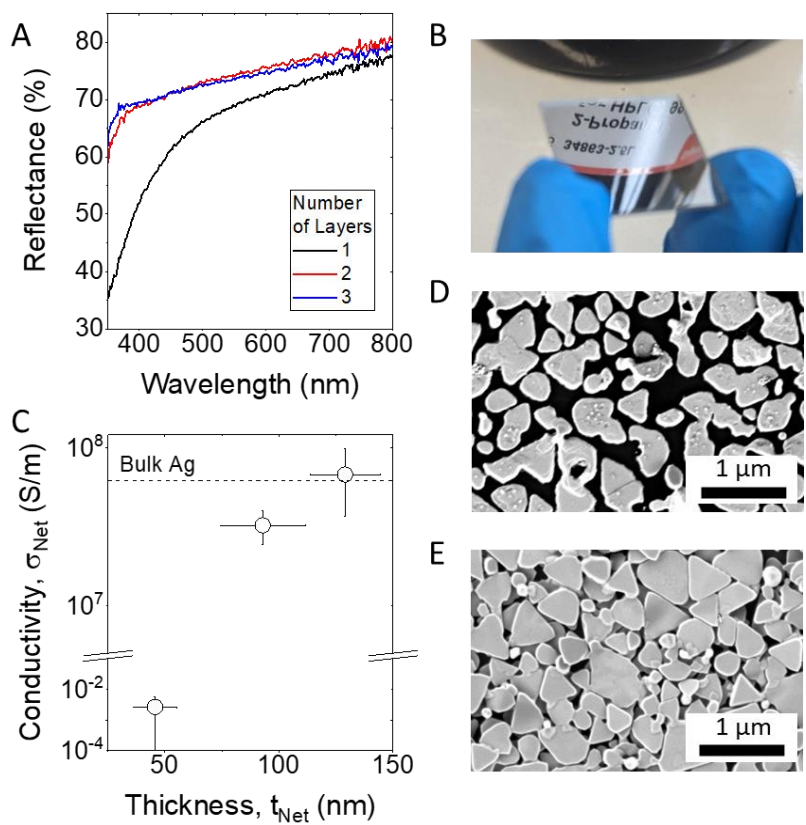

Fig 6. Beyond layered 2D materials. A. Reflectance spectra for 1L to 3L AgNS networks. B. Optical photograph of the Ag nanosheets monolayer on glass substrate, highlighting the specular reflectivity of the monolayer, such that the reflected text from a solvent bottle is clearly visible. C. Conductivity versus thickness plot for 1L to 3L AgNS networks. The y-axis has been split to accommodate the substantial conductivity range over 9 orders of magnitude. Annealed D. 1L and E. 2L AgNS network, showing the distinct morphology differences.